\documentclass[11pt]{article}
\newcommand{\Comment}[1]{{}}
\usepackage[textwidth = 430 pt, textheight = 630 pt]{geometry}
\usepackage{amssymb,euscript,amsmath,amsfonts}

\usepackage{color}
\definecolor{MyDarkBlue}{rgb}{0.15,0.15,0.45}
\usepackage[linktocpage=true]{hyperref}
\hypersetup{
colorlinks=true,
citecolor=MyDarkBlue,
linkcolor=MyDarkBlue,
urlcolor=MyDarkBlue,
pdfauthor={N.~Lambert, H.~Nastase and C.~Papageorgakis},
pdftitle={5D Yang-Mills instantons from ABJM Monopoles},
pdfsubject={hep-th}
}

\usepackage[numbers,sort&compress]{natbib}
\usepackage{hypernat}

\newcommand\ignore[1]{}
\def\one{{\,\hbox{1\kern-.8mm l}}}

\def\Tr{{\rm Tr\, }}

\newcommand{\SO}{\mathrm{SO}} 
\newcommand{\SU}{\mathrm{SU}} \newcommand{\U}{\mathrm{U}}
 \newcommand{\pd}{\partial}
\newcommand{\doublet}[2]{\left(\begin{array}{c}#1\\#2\end{array}\right)}

\newcommand{\Cset}{{\,\,{{{^{_{\pmb{\mid}}}}\kern-.45em{\mathrm C}}}}}

\newcommand{\nn}{\nonumber}
\newcommand{\ie}{{\it i.e.~}}
\newcommand{\eg}{{\it e.g.~}}
\newcommand{\be}{\begin{equation}}
\newcommand{\ee}{\end{equation}}
\newcommand{\bea}{\begin{eqnarray}}
\newcommand{\eea}{\end{eqnarray}}

\def\Rd{ R^{\dagger} }

\def\Gd{ G^{\dagger} }

\def\a{\alpha}

\def\d{\partial}
\def\dag{\dagger}

\def\Rd{ R^{\dagger} }

\begin{document}

\renewcommand{\thefootnote}{\fnsymbol{footnote}}

\rightline{CERN-PH-TH/2011-295}
\rightline{RUNHETC-2011-24}

   \vspace{1.8truecm}

\vspace{15pt}

\centerline{\LARGE \bf {\sc 5D Yang-Mills instantons from ABJM Monopoles}} \vspace{2truecm} \thispagestyle{empty} \centerline{
    {\large {\bf {\sc N.~Lambert,${}^{\,a,}$}}}\footnote{On leave of absence from King's College London.}$^,$\footnote{E-mail address: \href{mailto:neil.lambert@cern.ch}{\tt neil.lambert@cern.ch}} {}
    {\large {\bf{\sc H.~Nastase${}^{\,b,}$}}}\footnote{E-mail address:
                                 \href{mailto:nastase@ift.unesp.br}{\tt nastase@ift.unesp.br}} {and} {\large {\bf{\sc C.~Papageorgakis${}^{\,c,}$}}}\footnote{E-mail address:
                                 \href{mailto:papageorgakis@physics.rutgers.edu}{\tt papageorgakis@physics.rutgers.edu} }
                                                           }

\vspace{1cm}
\centerline{${}^a${\it Theory Division, CERN}}
\centerline{{\it 1211 Geneva 23, Switzerland}}
\vspace{.8cm}
\centerline{{\it ${}^a$ 
IFT, UNESP-Universidade Estadual Paulista}} 
\centerline{{\it S\~ao Paulo 01140-070, SP, Brazil}}
\vspace{.8cm}
\centerline{${}^c${\it NHETC and Department of Physics and Astronomy}}
\centerline{{\it Rutgers University, Piscataway, NJ 08854-8019, USA}}

\vspace{2.0truecm}

\thispagestyle{empty}

\centerline{\sc Abstract}

\vspace{0.4truecm}
\begin{center}
\begin{minipage}[c]{380pt}{
    \noindent In the presence of a background supergravity flux, $N$ M2-branes will expand via the Myers effect into M5-branes wrapped on a fuzzy three-sphere. In previous work the fluctuations of the M2-branes were shown to be described by the five-dimensional Yang-Mills gauge theory associated to D4-branes. We show that the ABJM prescription for eleven-dimensional momentum in terms of magnetic flux  lifts to an instanton flux of the effective five-dimensional Yang-Mills theory on the sphere, giving an M-theory interpretation for these instantons.}

\end{minipage}
\end{center}

\vspace{.4truecm}

\noindent

\vspace{.5cm}

\setcounter{page}{0}

\newpage
\renewcommand{\thefootnote}{\arabic{footnote}}
\setcounter{footnote}{0} 

\linespread{1.1}
\parskip 4pt

{}~
{}~

\makeatletter
\@addtoreset{equation}{section}
\makeatother
\renewcommand{\theequation}{\thesection.\arabic{equation}}

\section{Introduction}

The relation between the low-energy D4-brane theory and its corresponding M-theory counterpart, the low-energy M5-brane theory, poses some intriguing questions that are not  fully resolved. The conventional understanding is that, since five-dimensional maximally supersymmetric Yang-Mills (MSYM) is power-counting non-renormalisable, it is only a low-energy effective description which UV completes to the (2,0) theory on $S^1$.  In \cite{Douglas:2010iu,Lambert:2010iw} it was argued, following earlier leads by \cite{Rozali:1997cb,Berkooz:1997cq}, that BPS Kaluza-Klein states of the compactified six dimensional $(2,0)$ theory on $S^1$ can be recovered as instanton-charged states of five dimensional MSYM in flat space. This gave rise to the proposal that the latter is not in need of a UV completion in order to reproduce the finite six dimensional CFT at strong coupling and is a well defined QFT in its own right. This proposal has recently received further study and support from \cite{Bolognesi:2011nh,Bolognesi:2011rq,Tachikawa:2011ch,Kim:2011mv,Czech:2011dk,Young:2011aa}.

In particular, the instanton number of five-dimensional MSYM was identified with units of Kaluza-Klein momentum in the (2,0) theory on a circle of radius $R_5$ as
\be P_5=\frac{k}{R_5}=-\frac{1}{2g_{YM}^2}\int\Tr(F\wedge F)\label{kkinst}\;.
\ee
 Here we will see how this relation, and as a result the relation between the action and states of five-dimensional MSYM and the (2,0) theory, emerges in the context of the M2/M5 fuzzy sphere bound state arising from the Myers effect \cite{Myers:1999ps}. Some recent studies of M2/M5-brane systems include \cite{Terashima:2010ji,Gustavsson:2010ep}.

The M2/M5 system is interesting in this context, since it is expected to admit both an M2- and an M5-brane weakly-coupled description at large $N$. The former has been recently made accessible due to our enhanced understanding of M2-brane theories \cite{Bagger:2006sk,Gustavsson:2007vu,Bagger:2007jr,Bagger:2007vi,Aharony:2008ug} and is given in terms of the mass-deformed ABJM theory of \cite{Hosomichi:2008jb,Gomis:2008vc}. The latter should be captured by an M5-brane theory partially wrapping $S^3/\mathbb Z_k$. The first result of this paper is to fill a gap in the literature and explicitly construct the M5-brane picture by establishing this equivalence for large-$k$, where the M-theory circle shrinks in both descriptions. This is done by showing that the two theories share the same action for fluctuations. We carry this out in all detail for irreducible solutions of the mass-deformed ABJM theory and an Abelian M5-brane. We also give a concrete prescription for the extension of this matching between reducible ABJM solutions and multiple M5-branes on $S^3$.

We then incorporate momentum around the M-theory circle and show that the agreement still holds: From the ABJM point of view this involves turning on flux along the spatial worldvolume directions of the M2-brane theory. The associated charge is carried by the so-called monopole or 't Hooft operators. Our second result is that there exists a one-to-one map between this monopole charge and instanton charge in the five-dimensional  MSYM on $\mathbb R^{2,1}\times S^2$ both in the Abelian and non-Abelian cases. Note that instanton charge in five-dimensional MSYM is not only carried  by conventional self-dual gauge field configurations.

Our results imply that there is a one-to-one map between both perturbative and nonperturbative states of the mass-deformed ABJM and five-dimensional  MSYM theories. Since the former is the  complete description of the M2/M5-brane system \cite{Kim:2010mr,Cheon:2011gv}, correctly capturing momentum along the M-theory circle, the same should be true for the latter. This provides strong evidence for the conjecture of \cite{Douglas:2010iu,Lambert:2010iw}. We also hope that this paper helps to clarify the role of instanton number as Kaluza-Klein momentum. 

This result further suggests that instantons are important even in the Abelian theory. For example, we could wrap the M2-branes on a torus so that the massive vacua describe a single fuzzy M5 on $T^2\times S^3$. In this case the ABJM prescription for eleven-dimensional momentum maps to smooth   instantons in  the Abelian D4-brane picture. This is consistent with the conjectures of \cite{Douglas:2010iu,Lambert:2010iw} but yet apparently different from the standard prescription coming from the reduction of the known equations of motion for a single M5-brane \cite{Gustavsson:2011af,Linander:2011jy,Gustavsson:2011ur}. Thus there   appears to be  a duality between an Abelian D4-brane  on $T^2\times S^2$  with instantons and an Abelian M5 description on $T^2\times S^3$ including Kaluza-Klein modes of the Hopf fibration $S^3\stackrel{\pi}{\rightarrow} S^2$.

The rest of this article is organised as follows: In Section~\ref{setup} we consider the effective action of a single M5-brane in a four-form flux background. For states with no Kaluza-Klein momentum this is the same as a D4-brane in the associated type IIA background and we find that the brane is indeed stabilised at the correct radius as predicted by the gauge theory analysis. In Section~\ref{actionflucts} we then consider fluctuations of the D4-brane theory and show that this precisely agrees with the effective action calculated in \cite{Nastase:2009ny} from the ABJM description of M2-branes in the same background. These calculations are in precise agreement with the ABJM analysis and provide a non-trivial check on the dual description obtained from a D4-brane. In Section~\ref{nonabelian} we discuss the extension of this analysis to multiple D4-branes or M5-branes. For static configurations these can be obtained by considering reducible, block diagonal, representations of the ABJM vacuum equations.  Dynamically one must then also allow for off-diagonal modes whose effective action is given by Yang-Mills gauge theory. In Section~\ref{fluxes} we show that the ABJM prescription for momentum around the M-theory circle in terms of worldvolume $\U(1)$ flux is precisely mapped to the instanton number in the D4-brane description. We comment on the M-theory interpretation of this result. Finally Section~\ref{conclusions} contains our conclusions.

\section{Setup}\label{setup}

We want to consider an M5-brane probe in a particular flux background with\footnote{We work with conventions $A = \frac{1}{p!} A_{i_1\ldots i_p} dx^{i_1}\wedge \ldots \wedge dx^{i_p}$.}
\bea\label{flux}
  G^{(4)} &=&2\mu (dx^3\wedge dx^4\wedge dx^5\wedge dx^6 +dx^7\wedge dx^8\wedge dx^9\wedge dx^{10}) \ ,
\eea
where we note the factor of $2$. We observe that this solution is only at leading order in the fluxes. Due to gravitational effects there is a back-reaction on the geometry. The full solution is known \cite{Mohammed:2010eb,Lambert:2009qw}:
\bea
ds^2&=&H^{-2/3}(-dt^2+dx_1^2+dx_2^2)+H^{1/3}(dx_3^2+\ldots +dx_{10}^2)\\
\nonumber  G^{(4)} &=&2\mu (dx^3\wedge dx^4\wedge dx^5\wedge dx^6 +dx^7\wedge dx^8\wedge dx^9\wedge dx^{10}) +  d( H^{-1} -1) \wedge dt\wedge dx_1\wedge dx_2  \;,
\eea
where by directly solving the 4-form equation of motion $d\star G = \frac{1}{2}G \wedge G $ we get\footnote{Note that due to conventions we obtain a slightly different coefficient here than reported in \cite{Mohammed:2010eb}.}
\be
H =1 - \frac{1}{4}\mu^2 r^2\;.
\ee
 This solution is clearly singular when $H\sim 0$. Thus in order to trust the supergravity background we require that $\mu^2 r^2 \ll1$.

$N$ M2 probes were placed in  this flux background in   \cite{Lambert:2009qw}, leading to the derivation of the mass-deformed ABJM action of \cite{Hosomichi:2008jb,Gomis:2008vc}. Compared to the undeformed ABJM theory \cite{Aharony:2008ug} there is a correction to the supersymmetry transformations of the M2-branes due to the flux, given by 
\be
\delta \psi_A = \gamma^\mu D_\mu Z^B\epsilon_{AB} + [Z^C,Z^D;\bar Z_A]\epsilon_{CD}
+[Z^D,Z^C;\bar Z_D]\epsilon_{AC} + \frac{1}{2}M_A{}^CZ^D\epsilon_{CD}\;,
\ee
where $A = 1,...,4$, $[Z^A,Z^B;\bar Z_C] \equiv \frac{2\pi}{k}(Z^A Z_C^\dag Z^B - Z^BZ_C^\dag Z^A)$ and
\be
M_A{}^B = 2\mu\left(
\begin{array}{cccc}
1 &   &    &    \\
  & 1 &    &    \\
  &   & -1 &    \\
  &   &    & -1 \\
\end{array}\right)\;.
\ee
Setting $Z^3=Z^4=0$ leads to the vacuum equation 
\be\label{BPSvac}
[Z^A,Z^B;\bar Z_B] = \mu Z^A \qquad A,B=1,2\ ,
\ee
which is the one that was  used in \eg \cite{Gomis:2008vc,Nastase:2009ny}, justifying the factor of $2$ that appears in (\ref{flux}). In the presence of this flux, the M2's will exhibit a multi-pole coupling to  M5-brane charge through terms $\int d^3x\;  C^{(6)} \Tr(D\bar Z[Z,Z;\bar Z]) + h.c.$ in an M-theory realisation of the Myers effect \cite{Myers:1999ps,Bena:2000zb,Lambert:2009qw}. The resulting configuration is an M2/M5 bound state where the M2's have been blown up into an $S^3$ inside one of the ${\mathbb R}^4$ sub-planes of the transverse eight dimensional space. 

For large $N$ this M2/M5 bound state is expected to  have an equivalent M5-brane description in terms of a spherical M5 in the same four-form flux background with an additional worldvolume self-dual flux $H$. The latter prevents the sphere from collapsing under its own tension. The system carries M2-brane charge through the coupling $\int   C^{(3)}\wedge H$. Moreover, since in the  ABJM description we have to  consider the branes on a $\mathbb Z_k$ orbifold singularity, we  also wish to think of the background, in the absence of flux, as ${\mathbb R}^8/\mathbb Z_k$ or $({\mathbb R}^4\times {\mathbb R}^4)/\mathbb Z_k$ and then factor out a common $\U(1)$ fibre where the $\mathbb Z_k$ acts.

 In particular, we consider `spherical' coordinates for ${\mathbb R}^4$ and use the standard Hopf fibration of the unit $S^3$
\be
ds^2_{S^3} = \frac{1}{4}\left(d\theta^2+\sin^2\theta d \phi^2 +  (d\psi +\cos\theta d\phi)^2\right)\;,
\ee
with   $\psi  \in [0,4\pi)$, $\theta  \in [0,\pi]$ and $\phi \in [0,2\pi)$.
Thus
we proceed by considering the background geometry  \cite{Aki}
\bea\label{geometry}
\nonumber ds^2&=&H^{-2/3}(-dt^2+dx_1^2+dx_2^2)+H^{1/3}(d\vec{r}_1^2+ r_1^2(d\psi_1+A_1)^2+d\vec{r}_2^2+ r_2^2(d\psi_2+A_2)^2)\\
d\vec{r}_i^2&=&dr_i^2+\frac{1}{4}r_i^2(d\theta_i^2+\sin^2\theta_i d\phi_i^2)\cr
A_i&=&\cos\theta_i \frac{d\phi_i}{2} \;.
\eea
Note that the 2-spheres   have radius $\frac{1}{2}$ and we have redefined $\psi_i$ to have period $2\pi$.
We expect the M5-brane to blow-up into a 3-sphere with coordinates $\psi,\theta,\phi$ and unit radius. 

Further defining
\bea
\psi_1&=&\tilde \psi+\psi\cr
\psi_2&=&\tilde \psi\ ,
\eea
we want to  implement the orbifold identification on $\tilde \psi$, so as to have $\tilde \psi\sim \tilde \psi+\frac{2\pi}{k}$, and then to dimensionally 
reduce on that direction. To this end we introduce the variable
\be
x_{11}= {kR_*} \tilde\psi\;,
\ee
so that $x_{11}$ has dimensions of length and periodicity $2 \pi R_{*}$. We will use this coordinate every time we reduce down to ten dimensions.  

One can express
\bea
r_1^2(d\psi_1+A_1)^2+r_2^2(d\psi_2+A_2)^2&=&\frac{r_1^2r_2^2}{r_1^2+r_2^2}(d\psi+A_1-A_2)^2+(r_1^2+r_2^2)(d\tilde \psi+{\cal A})^2\cr
{\cal A}&=&\frac{r_2^2A_2+(d\psi+A_1)r_1^2}{r_1^2+r_2^2}\;.
\eea
Then, through the general reduction formula
\be
ds_{11}^2=e^{-2\varPhi/3}ds_{10}^2+e^{4\varPhi/3}(dx_{11}+C^{(1)})^2 \;,
\ee
we obtain
\bea\label{metric}
ds_{10}^2&=&\frac{\sqrt{r_1^2+r_2^2}}{kR_{*}}\Big(H^{-1/2}(-dt^2+dx_1^2+dx_2^2) \cr
&&\qquad\qquad\qquad\qquad\qquad+ H^{1/2}\Big[d\vec{r}_1^2+d\vec{r}_2^2+\frac{r_1^2r_2^2}{r_1^2+r_2^2}(d\psi+A_1-A_2)^2\Big]\Big)\cr
e^\varPhi&=&\Big( \frac{\sqrt{ r_1^2+r_2^2}}{kR_{*}}\Big)^{3/2}H^{1/4}\cr
C^{(1)}&=& {kR_{*}} {\cal A}\;.
\eea

We next consider the reduction of the 11d flux:
\bea
\nonumber G^{(4)} &=&2\mu (dx^3\wedge dx^4\wedge dx^5\wedge dx^6 +dx^7\wedge dx^8\wedge dx^9\wedge dx^{10})+dt\wedge dx_1\wedge dx_2\wedge dH^{-1}\\
   &=&2\mu  \sum_{i=1,2}r_i^3\sin\theta_i dr_i \wedge d\theta_i\wedge d\phi_i\wedge  d\psi_i + dt\wedge dx_1\wedge dx_2\wedge dH^{-1}\;.
\eea
From this we can compute
\begin{equation}\label{CS3}
C^{(3)}= \frac{\mu}{2}\sum_{i=1,2}r_i^4\sin\theta_id\theta_i\wedge d\phi_i\wedge d\psi_i
+(H^{-1}-1)dt\wedge dx_1\wedge dx_2\ ,
\end{equation}
so that the ten-dimensional 2-form is
\begin{equation}
B = \frac{\mu }{2kR_*}[r_1^4\sin\theta_1d\theta_1\wedge d\phi_1+r_2^4\sin\theta_2 d\theta_2\wedge d\phi_2]\;.
\end{equation}
We also need to look at
\bea
\nonumber G^{(7)} &=& dC^{(6)}\\
&=& \star G^{(4)} - \frac{1}{2} C^{(3)}\wedge G^{(4)}\\
\nonumber &=&2\mu H^{-1}dt\wedge dx_1\wedge dx_2\wedge \sum_{i=1,2}r_i^3\sin\theta_i dr_i\wedge d\theta_i \wedge d\phi_i \wedge d\psi_i +\ldots\;,
\eea
where in the above the ellipsis denotes terms which will not be needed in the rest of the calculation, \eg terms with no $dt\wedge dx^1\wedge dx^2$ factor. Thus in ten dimensions we can effectively use
\be\label{asymf}
C^{(5)} =- \Big(\frac{\mu}{  4kR_*} H^{-1}+ \frac{\mu}{4k R_*}\Big)   dt\wedge dx_1\wedge dx_2 \wedge \sum_{i=1,2}r_i^4\sin\theta_i  d\theta_i \wedge d\phi_2+\ldots \;.
\ee

\subsection{Brane Embedding}

We now wish to introduce a brane probe into this background. In particular, we can consider either an M5-brane in the eleven-dimensional spacetime or, as in this section we are looking at static solutions where there is a $\U(1)$ isometry, a D4-brane in the ten-dimensional spacetime. Since the action for a D4-brane is unambiguous we chose to work with the latter.  The D4-brane should extend in the $t,x_1,x_2$ directions and two more directions along an $S^2$. For definiteness, we will choose those to be $\theta_1,\phi_1$.

In addition, we would like to have a net D2-brane charge of $N$ units in the system. As a result we require the presence of a background worldvolume flux
\be
{\cal F}_0 = \lambda F_0 - P[B] = \Big( {2\lambda N} -\frac{ \mu r_1^4}{ 2kR_*}\Big)\sin\theta_1 d\theta_1\wedge d\phi_1\ ,\label{backflux}
\ee
with ${F_0}_{\theta_1 \phi_1} = 2 N \sin{\theta_1}$, so that\footnote{Recall that the two-sphere has radius $\frac{1}{2}$.}
\be
T_{D4}\int _{{\mathbb R}^3\times S^2}C^{(3)}\wedge ({F_0}-P[B]) = N T_{D2}\int_{{\mathbb R}^3}C_{012} dt\wedge dx_1\wedge dx_2 + \ldots\;,
\ee
where $NT_{D2} = 4\pi^2 \alpha' N T_{D4}$. In the definition of $\mathcal F_0$ we have taken into account the pullback of the background $B$-field, although its contribution to the D2-brane charge will turn out to be subleading upon finding the vacuum of the theory. One can check that ${F}_0 $ satisfies the D4-brane equations of motion in a vacuum where the scalar fields are constant.

With the above in mind, the general form for the effective action is
\be\label{effective}
S = -T_{D4}\int d^5x \; e^{-\varPhi}\sqrt{-\det(g+\cal F)} -T_{D4}\int \Big[P[C^{(5)}]+P[C^{(3)}]\wedge {\cal F} +\frac{1}{2}P[C^{(1)}]\wedge 
{\cal F}\wedge {\cal F}\Big]\;,
\ee
where ${\cal F}_{\mu\nu} =  (\lambda F_{\mu\nu} - P[B]_{\mu\nu})$, with $\lambda = 2\pi\alpha'$, and $g$ is the pullback of the spacetime metric. 

Let us now determine how the D4-brane is embedded.  We will write
\bea
&&r_1=\rho \cos\xi\cr
&&r_2=\rho \sin\xi
\eea
and suppose that the brane is at some $ \rho = R$. 
We then proceed to evaluate the effective action \eqref{effective}. The DBI term simply comes out of using the metric in (\ref{metric}) and the flux in (\ref{flux}), while the CS terms come from the 
$C_{012\theta_1\phi_1}$ term in (\ref{asymf}) and $C_{012}$ terms in \eqref{CS3}. Then, ignoring fluctuations, we obtain
\be\label{sqrt}
\begin{split}
 S = -T_{D4} \cos^2{\xi}\int d^5x\sin\theta_1  &\left( H^{-1/2}\frac{ R^3}{  kR_{*}}\sqrt{1+H^{-1}\frac{ k^2R_{*}^2}{R^6  \cos^4\xi}\Big( {2\lambda N}   - \frac{\mu R^4}{2 kR_*}\Big)^2}\right. \\ & \left. + \frac{2 \lambda N (H^{-1}-1)}{\cos^2{\xi}}- \frac{ \mu R^4}{4kR_{*}}(H^{-1}+1) \cos^2\xi \right) \;.
\end{split}
\ee

We first note that minimising with respect to $\xi$ we get $\sin\xi=0\Rightarrow\cos\xi=1$, \ie we are actually in the $R=r_1$ case. As a result in what follows we will simply set $r_2=0$.\footnote{For the choice of the embedding of the D4 in the $\theta_2, \phi_2$  sphere, one would have obtained $\sin\xi=1$ and as a result  $R = r_2$.}

Let us now expand the square root of \eqref{sqrt}. In the large $N$ limit, {\it i.e.} to leading order in $R^3/\lambda N k R_*$, we have
\bea\label{expanded}
\nonumber S &=& -T_{D4} \int d^5x \sin\theta_1 \left( {2\lambda N} + {\lambda N} \mu^2 R^2+\frac{ R^6}{4\lambda Nk^2R_{*}^2} - \frac{  \mu R^4}{kR_{*}}  \right)\\
&=& -T_{D4} \int d^5x \sin\theta_1 \left( {2\lambda N} + {\lambda N} \mu^2 R^2\left(1-\frac{ R^2}{2\lambda \mu NkR_{*}}\right)^2  \right)\\
\nonumber  &=& -T_{D4} \int d^5x \sin\theta_1 V(R)\;.
\eea
Note that here we have also expanded $H^{-1} = 1 +\frac{1}{4}\mu^2 R^2$ by assuming the approximation $\mu^2R^2\ll 1$ that was needed to ensure the validity of the supergravity solution.

Clearly $V(R)$ has stable vacuum solutions corresponding to radii
\be\label{extremum}
R_0 = 0\qquad\textrm{and}\qquad R_0^2 = 2\mu \lambda k N R_*\;.
\ee
There is also a local maximum at $R_0^2 = \frac{2}{3}\mu \lambda k N R_*$ which we discard. 
Our approximation $R^3/\lambda N k R_*\ll 1$, used in the square root expansion above, now  becomes
\be
R_0 \mu\ll 1 \;.
\ee

Note also that by construction $2\pi R_*$ is the periodicity of the eleventh dimension; $x^{11}\cong x^{11}+2\pi R_*$. Thus following the usual M-theory/Type IIA relations 
one can identify $R_* = g_sl_s$ and $l_p = g_s^{1/3} l_s$ where $g_s = e^{\langle\varPhi \rangle}$ and $l_s=\sqrt{\alpha'}$. Therefore we find 
\be
R_0^2 = 4{\pi} \mu k l_p^3 N
\ee
and this agrees with the result from the M2-brane description   given by Eq.~(7.7) in \cite{Nastase:2009ny}.
  
Finally, if we make the choice $R_* = \frac{R_0}{k}$, so that $R_*$ is also the radius of the M-theory circle as measured at $R_0$ in the eleven-dimensional metric,  then  we arrive at 
\be\label{minimum}
R_0 = 2 \lambda  \mu N\;.
\ee
This is exactly the value for the physical radius evaluated in Eq.~(7.10) of \cite{Nastase:2009ny} and corresponds to an M-theory configuration where the M5-brane is wrapping a fuzzy $S^3$ realised as the Hopf fibration
\be
S^1/\mathbb Z_k \hookrightarrow S^3_F/\mathbb Z_k \stackrel{\pi}{\rightarrow} S_F^2\;.
\ee

\section{The action for fluctuations}\label{actionflucts}

We next study fluctuations around the above solution for the bosonic fields. We first focus on the gauge field and radial scalar and then move on to the remaining scalar fields.

\subsection{Gauge fields and radial scalar}

Consider fluctuations  with only $\partial_\mu \delta R$ and $\delta {\cal F}$ non-vanishing. The action has the form\footnote{The $P[C^{(1)}]\wedge \mathcal F \wedge \mathcal   F$ term of \eqref{effective} will result in subleading contributions and will hence be ignored.}
\bea
S&=& -T_{D4} \int d^5x e^{-\varPhi} \sqrt{-\det(G_0+\Upsilon) }-T_{D4}\int d^5 x (C^{(5)}_{012\theta_1\phi_1}+C^{(3)}_{012}{\cal F}_{\theta_1
\phi_1})\;,
\eea
where
\bea
\nonumber \Upsilon_{\mu\nu}&=& \partial_{\mu}\delta R\partial_{\nu} \delta R g_{rr} + \lambda\delta { F}_{\mu\nu}\\
G_{0\mu\nu} &=& g_{\mu\nu} + {\cal F}_{0\mu\nu}\;.
\eea
Expanding to quadratic order in $\Upsilon$ we find
\bea
\nonumber S&=& -T_{D4} \int d^5x e^{-\varPhi} \sqrt{-\det G_0}\left(1 +\frac{1}{2}\Tr (G_0^{-1}\Upsilon)-\frac{1}{4} \Tr ((G_0^{-1}\Upsilon)^2) +\frac{1}{8}( \Tr( G_0^{-1}\Upsilon))^2\right) \\
&&-T_{D4}\int d^5 x (C^{(5)}_{012\theta_1\phi_1})\;.
\eea

Note that  $G_0$ can be evaluated from the previous section and is valid for arbitrary but constant values of $R$.
Explicitly we find
\be
G_0 =  \frac{ RH^{-1/2}}{kR_*}
\left(
\begin{array}{ccccc}
- 1 &   &  & & \\
  & 1  &   &  & \\
  &  &1 & &    \\
  & && HR^2&\Sigma{\sin\theta_1} \\
  & & &-\Sigma{\sin\theta_1}  & H R^2\sin^2\theta_1
\end{array}
\right)\ ,
\ee
where 
\be
\Sigma = \frac{kR_*}{ R}H^{ 1/2}(2 {\lambda N} - \frac{ \mu R^4}{ 2kR_*}) \ .
\ee
Therefore
 \be
G^{-1}_0 =  \frac{kR_*}{ RH^{ -1/2}}
\left(
\begin{array}{ccccc}
- 1 &   &  & & \\
  & 1  &   &  & \\
  &  &1 & &    \\
  & && \Delta^{-1}H^2R^4\sin^2\theta_1 &- \Delta^{-1}\Sigma{\sin\theta_1} \\
  & & &   \Delta^{-1}\Sigma {\sin\theta_1} & \Delta^{-1}H R^2
\end{array}
\right)\ ,
\ee
where
\be
\Delta  =   H \frac{k^2R_*^2}{ R^2}\Big(2{\lambda N} - \frac{\mu R^4}{2kR_*}\Big)^2 \sin^2\theta_1 \ .
\ee
Finally, as before, we have
\be 
e^{-\varPhi} \sqrt{-\det G_0} = \sin\theta_1\sqrt{H^{-1}\frac{ R^6}{k^2R_*^2}+H^{-2}\Big(2{\lambda N} - \frac{ \mu R^4}{2kR_*} \Big)^2}\ .
\ee

In order to proceed, it will be useful to split
\be
G_0^{-1} = D + A\;,
\ee
with $D$ diagonal and $A$ antisymmetric. The action for fluctuations can then be written as
\bea
\nonumber S&=& -T_{D4} \int d^5x e^{-\varPhi} \sqrt{-\det G_0}\Big(1 +\frac{1}{2} \Tr (D\partial\delta R\partial \delta R g_{r_1r_1}  )\\
\nonumber && +\frac{1}{2}\Tr (A\delta {\cal F}) +\frac{1}{8}[\Tr(A\delta { \cal   F})]^2- \frac{1}{4} \Tr[ (D\delta {  \cal  F})^2 + (A\delta {  \cal  F})^2 ]\Big) \\
&&-T_{D4}\int d^5 x (C^{(5)}_{012\theta_1\phi_1})\;.
\eea

First, we see that there is a linear term in $\delta {\cal F} = \lambda\delta F$:
\be
\frac{1}{2}e^{-\varPhi} \sqrt{-\det G_0}\Tr (A \delta {  F}) =   \frac{1-\frac{ \mu R^4}{4k\lambda NR_*}}{\sqrt{\left(1-\frac{ \mu R^4}{4k\lambda NR_*}\right)^2 + \frac{H  R^6}{4\lambda^2 N^2k^2R_*^2}}} H^{-1} \lambda\delta F_{\theta_1\phi_1}\;.
\ee
To evaluate the above we note that it has the form
\bea
\frac{1-X}{\sqrt{(1-X)^2+(1-Z)Y}}\frac{1}{1-Z}\lambda \delta F_{\theta_1\phi_1}\;,
\eea
where
\be
X = \frac{ \mu R^4}{4k\lambda NR_*}\;, \qquad Y=\frac{ R^6}{4\lambda^2 N^2k^2R_*^2}\;, \qquad Z =\frac{1}{2}\mu^2 R^2\;.
\ee
Now in our approximation $X,Y,Z\ll 1$ but all are of the same order (at $R=R_0$). Expanding $R=R_0+\delta R$ gives to leading order 
\bea
\nonumber \frac{1}{2}e^{-\varPhi} \sqrt{-\det G_0}\Tr (A \delta {  F}) &=&\Big(1-\frac{1}{2}Y+Z\Big)\lambda \delta F_{\theta_1\phi_1}\\
&=&  \lambda\delta F_{\theta_1\phi_1}+\lambda\left( -\frac{6R_0^5}{8\lambda^2 N^2 k^2 R_*^2}+\mu^2 R_0\right)  \delta R\delta F_{\theta_1\phi_1}\\
\nonumber &=& \lambda\delta F_{\theta_1\phi_1} - 2\lambda \mu^2 R_0  \delta R\delta F_{\theta_1\phi_1}\;.
\eea
The first term in the last line is a total derivative and  can be discarded, which is compatible with the fact that the gauge field is on-shell.

We also need to look at the terms quadratic in $\delta F$. The quadratic terms involving the anti-symmetric part $A$ cancel. In addition we note that to leading order, and for terms that involve quadratic fluctuations,  we can simply take
\bea
 e^{-\varPhi} \sqrt{-\det G_0} &=& {2\lambda N}   \nonumber\\
\nonumber H&=&1\\
\Delta &=& \frac{4\lambda^2N^2k^2R_*^2}{ R ^2} \sin^2\theta_1 \\
\nonumber D &=& \frac{kR_*}{ R}
\left(
\begin{array}{ccccc}
-  1
 &   &  & & \\
  &  1
  &    &   & \\
  &  & 1
 & &    \\
  & & &   \frac{ R^4}{4\lambda^2 N^2k^2R_*^2}&  \\
  & & &  &\frac{ R^4}{4\lambda^2 N^2k^2R_*^2 \sin^2\theta_1}
\end{array}
\right)  \;.
\eea
 Putting these together we find
 \be\label{38}
\begin{split}
 S =  -T_{D4} \int d^3x d\theta_1 d\phi_1  \sin\theta_1& \Big[ V(R) + 
 {\lambda N} \Tr (D\partial\delta R\partial \delta R g_{r_1r_1}  )\\
& -\frac{2\lambda\mu^2R_0}{\sin\theta_1}\delta R \delta F_{\theta_1\phi_1} -\frac{\lambda^3 N}{2} \Tr[ (D\delta {  F})^2  ]
  \Big]\;.
\end{split}
\ee

Next we need to expand $R =R_0+\delta R$ in the potential $V(R)$.  
\be
V(R) = 2{\lambda N} + 4\mu^2\lambda N (\delta R)^2\;.
\ee

Combining all terms we arrive at
\bea\label{44}
\nonumber S&=& -\frac{1}{4}T_{D4} \mu^{ 2} \int d^5x \sqrt{\det h} \Big[ 2 {\lambda N}   +  {4\mu^2\lambda N}  (\delta R)^2+  
 {\lambda N} \partial_\mu\delta R\partial^\mu \delta R \\
&& -\frac{2 \lambda\mu^2 R_0}{\sin\theta_1}   \delta F_{\theta_1\phi_1}\delta R+\frac{\lambda^2kR_*}{4\mu }\delta F_{\mu\nu}\delta F^{\mu\nu}
  \Big]\nn\\
\nonumber &=& -\frac{2\lambda N}{4}T_{D4} \mu^{ 2} \int d^5x \sqrt{\det h} \Big[ 1 +
 \frac{1}{2}\partial_\mu\delta R\partial^\mu \delta R \\
&&+\frac{1}{4}\left(\lambda \delta F_{\mu\nu}-  2  \mu \delta R\omega_{\mu\nu}  \right) \left(\lambda \delta F^{\mu\nu}- 2 \mu \delta R \omega^{\mu\nu} \right) 
  \Big]\;.
\eea
In the above the indices are raised and lowered with the metric on a spacetime ${\mathbb R}^{1,2}\times S^2$, where the sphere has metric $h$,  radius $\mu^{-1}$ and $\mu,\nu=\{0,1,2,\theta_1,\phi_1\}$.\footnote{The fact that the $S^2$ has radius $\mu^{-1}$ instead of $\frac{1}{2}$ is achieved by a scaling of the sphere metric.} In the last line we also used $R_0 = k R_*$  and have introduced the symplectic form on the sphere
\be
\omega = \sqrt{\det h}\; d\theta_1\wedge d\phi_1\ .
\ee

Finally, by  expressing the scalar field in terms of $\delta R = \lambda \delta \Phi$, the 
 action comes to the familiar  form\footnote{We have dropped the constant term.}
\be\label{final}
S = \frac{1}{g^2}\int d^5 x\sqrt{\det h} \Big[
 \frac{1}{2}\partial_\mu\delta \Phi\partial^\mu \delta \Phi+\frac{1}{4}\left(\delta F_{\mu\nu}-  2  \mu \delta \Phi\omega_{\mu\nu}  \right) \left(\delta F^{\mu\nu}- 2 \mu \delta \Phi \omega^{\mu\nu} \right)\Big]\;,
\ee
where $g^2 = { 4 g_s l_s (2\pi)^2}/{R_0 \mu}$.\footnote{The overall coupling can always be changed by a further simultaneous rescaling of all fields, since the action is quadratic.}

\subsection{Fluctuation Analysis}\label{analysis}

A few comments are in order: First, we note that  Eq.~\eqref{44} is the same result as the one for the action of fluctuations in the ABJM calculation of \cite{Nastase:2009ny}. In particular, it is useful to compare the coefficient of the $\delta F_{\mu\nu}\delta F^{\mu\nu}$ term between the `M2' and `D4' calculations. Examining Eq.~(6.1) of \cite{Nastase:2009ny} we see that
\be\label{them}
S_{M2} = -\frac{1}{4\pi}\frac{k\mu }{16\pi} \int d^5 x\sqrt{\det h}F_{\mu\nu}  F^{\mu\nu}+\ldots\;.
\ee
Note the additional coefficient of $\frac{1}{4\pi}$ compared to \cite{Nastase:2009ny}. This arises because, when switching from matrices to geometry in the large-$N$ limit, one should make the identification
\be
\frac{1}{N}{\rm Tr} \longrightarrow \frac{1}{4\pi}\int_{S^2} d^2x\sqrt{\det \hat h}\ ,
\ee
\ie including the normalisation $\frac{1}{4\pi}$, where $\hat h$ is the metric on the unit  2-sphere while $h$ is the metric on the sphere of radius $\mu^{-1}$ that we are interested in. 
We observe that $T_{D4}=2\pi R_* T_{M5} = R_*T_{M2}^2 = (2\pi)^{-4}l_p^{-6}R_*$. Noting that  $R_* = g_sl_s$ and $l_p^3=g_sl_s^3$ we find 
$T_{D4}\lambda^2 R_* = (2\pi)^{-2}$. Then we have
\be
S_{D4} = -\frac{k\mu}{64\pi^2 } \int d^5 x\sqrt{\det h}\delta F_{\mu\nu} \delta F^{\mu\nu}+\ldots\;,
\ee
which agrees exactly with (\ref{them}) upon identification of $F_{M2} \equiv \delta F_{D4}$.

 Thus the effective action obtained from examining fluctuations of M2-branes about a mass-deformed vacuum using the ABJM description precisely agrees with that obtained from a single D4-brane in type IIA in the same flux background.  In particular, the components of the D4-brane gauge field along ${\mathbb R}^{2,1}$ can be identified with the overall $\U(1)$ of the $\U(N)$ diagonal subgroup coming from the Higgsing of the $\U(N)\times\U(N)$ ABJM gauge fields \cite{Nastase:2009ny, Nastase:2009zu,Nastase:2010uy}.

We will now argue that the above agreement is still valid in the case where the vacuum we expand around involves a nonzero constant $F_{12}$ worldvolume flux. Clearly such a configuration is a vacuum solution to the mass-deformed ABJM equations of motion if we note the following fact: Turning on an equal $\U(1)$ background flux for the left and right ABJM gauge fields does not have any effect on the dynamics, since in this case
\be
D_\mu Z^A = \pd_\mu Z^A - i (A_\mu^L - A_\mu^R) Z^A = \pd_\mu Z^A ,
\ee
and hence this particular flux does not couple to the matter fields. Hence one can turn on $F_{12}$ without modifying  the calculation for the action of fluctuations already present in \cite{Nastase:2009ny, Nastase:2009zu,Nastase:2010uy}. Nevertheless this flux is important and corresponds to turning on momentum in the M-theory picture, in a fashion that we will describe in Section~\ref{fluxes}. 

Second, it might be surprising at first that the sphere metric appearing in \eqref{44} is not part of the pull-back metric on the D4-brane. However, this is not unusual: In the context of open string excitations in the presence of a closed string background with a B-field the open and closed string modes see a different metric. There is also an induced noncommutativity on the worldvolume theory, controlled by the parameter $\varTheta$ \cite{Seiberg:1999vs}. This has  been observed beyond the flat brane case for the D0-D2 dielectric configuration in \cite{Papageorgakis:2005xr}. 

The open string metric and noncommutativity parameter are given by
\be\label{openclosed}
{G_{(open)}}^{\mu\nu} = \Big(\frac{1}{G_{(closed)} + \lambda F_0} \Big)^{\mu\nu}_\textrm{symmetric}\quad \textrm{and}\quad \varTheta^{\mu\nu} = \Big(\frac{1}{G_{(closed)} + \lambda F_0} \Big)^{\mu\nu}_\textrm{antisymmetric}\;,
\ee
which are precisely our definitions of $D$ and $A$ respectively. Hence, up to the conformal factor $\frac{k R_*}{R_0}$ due to the nontrivial dilaton, and once again in the limit where we make use of $\mu^2 R_0^2\ll 1$, this is exactly what our fluctuations see. In particular, for the natural choice which gives agreement with \cite{Nastase:2009ny}, $R_* = \frac{R_0}{k}$, the dilaton factor drops out and we have 
\be
h^{-1} =
\left(
\begin{array}{ccccc}
-  1
 &   &  & & \\
  &  1
  &   &  & \\
  &  & 1  & &    \\
  & & & \mu^2 &  \\
  & & &  &\mu^2\frac{1}{\sin^2\theta_1}
\end{array}
\right) \quad\textrm{and}\quad 
\varTheta =
\left(
\begin{array}{ccccc}
  0 &   &  & & \\
  & 0  &   &  & \\
  &  &0 & &    \\
  & & & 0& -\frac{ 1}{2\lambda N \sin{\theta_1}}\\
  & & & \frac{1}{2\lambda N \sin{\theta_1}} &0
\end{array}
\right)\;.
\ee
It is obvious that in our large-$N$ limit the noncommutativity parameter vanishes and the resulting theory is an ordinary $\U(1)$ gauge theory. However, its existence is important if the D4-brane action is to reproduce the fuzzy sphere geometry at finite $N$.

Finally, we note that the equation for a constant $\delta \Phi$ (VEV) is 
\be
\mu \delta \Phi=\frac{1}{4}{\omega^{ab}\delta F_{ab}}
\ee
where $a,b=\theta_1,\phi_1$. On the other hand the equation for $\delta A_a$ gives 
\be
\delta F_{ab}=2\mu\delta \Phi\omega_{ab}+ G_{ab}\;,
\ee
where $G_{ab}$ satisfies $\d^a   G_{ab}=0$ and $\omega^{ab}  G_{ab}=0$. This has the only solution $G_{ab}=0$ and leads to a vanishing on-shell action. In the absence of the $\mu$ deformation, the action $-\frac{1}{4}\delta F_{ab}^2$ would allow for constant flux solutions,  $\delta F_{ab}=c\omega_{ab}$. However, these are not allowed in the case at hand, since $\delta \Phi=0$ is not a consistent truncation. Of course, we can still have constant flux solutions in the $x^0,x^1,x^2$ directions.

In particular note that the action has an infinite class of vacuum solutions:
\be\label{VacSol}
\delta F_{ab}= \frac{n}{2} \omega_{ab}\ ,\qquad \delta \Phi = \frac{n}{4\mu}
\ee
with vanishing action and quantised flux $n$ through $S^2$. However these solutions correspond to changing the number of M2-branes in the background by $n$ and also the value of the stabilised radius. As such we should not consider them as valid solutions of the effective theory with the boundary conditions we have imposed (namely that there are a fixed number of M2-branes). Indeed such solutions cannot arise if we assume that $\delta A_a$ is globally defined, `small' fluctuation. On the other hand we will see that allowing for magnetic flux $\delta F_{12}$ through the non-compact spatial dimensions  plays the physical role of introducing eleven-dimensional momentum into the effective theory.

\subsection{Overall Transverse Scalars}\label{overalltransverse}

We now turn our attention to the overall transverse scalars,  that is fluctuations in the directions transverse to both the D4-brane and the radius of the sphere. In order to study these fluctuations we revisit the expression for the ten-dimensional metric from \eqref{metric}. In the limit $\xi\to 0$ and $H=1$ this can be approximated by
\be\begin{split}
ds_{10}^2= &\frac{\rho}{k R_{*}}\Big(-dt^2+dx_1^2+dx_2^2+d\rho^2+ \rho^2 (d\theta_1^2+\sin^2\theta_1d\phi_1^2)\Big)\\
&+\frac{\rho^3}{kR_{*}} \Big( d\xi^2 + \xi^2(d\theta_2^2+\sin^2\theta_2d\phi_2^2)
+\xi^2(d\psi+A_1-A_2)^2\Big)\\
\simeq&\frac{\rho}{kR_{*}}\Big(-dt^2+dx_1^2+dx_2^2+d\rho^2+ \rho^2 (d\theta_1^2+\sin^2\theta_1d\phi_1^2)\Big)\\
&+ dX_6^2 + dX_7^2 + dX_8^2 + dX^2_9
\end{split}\;,
\ee
since the second line of the above essentially describes the origin of $\mathbb R^4$ in an $S^3$ foliation.
 
Alternatively, one could have started with \eqref{geometry} in the limit where $r_2\to 0$. In this limit one is fixed at the origin of the second $\mathbb R^4$ factor (or $\mathbb C^2$ parametrised by $Z^{\dot\alpha}$, where $\dot\alpha = 1,2$) and as such the $\mathbb Z_k$ orbifold projection, with $Z^{\dot \alpha} \to Z^{\dot \alpha} e^{\frac{2\pi i}{k}} = Z^{\dot \alpha} (1 + \frac{2\pi i }{k}+ \ldots)$ for large $k$, does not have an effect on the fluctuations in these directions:
\be
 Z^{\dot \alpha}\to (0 + \delta Z^{\dot \alpha}) \left(1 + \frac{2\pi i }{k} + \ldots\right)\simeq \delta Z^{\dot \alpha} + \ldots\;.
\ee

In summary, one could have started in the approximation where the background geometry in the absence of fluxes is $\mathbb R^{2,1}\times \mathbb R^4/\mathbb Z_k \times \mathbb R^4$, with $ \mathbb R^4/\mathbb Z_k$ realised in terms of an $S^1/{\mathbb   Z}_k\hookrightarrow S^3/{\mathbb Z}_k\stackrel{\pi}{\rightarrow} S^2$ foliation, to obtain exactly the same results for the D4-brane effective action and the action for fluctuations.

Making use of the above, it is straightforward to  modify our equations and include the  fluctuations of the transverse scalars. One has that 
\be
\Upsilon_{\mu\nu} = \pd_\mu \delta R \pd_\nu \delta R g_{rr}+ \lambda \delta F_{\mu\nu} + \pd_\mu  \delta X^m \pd_\nu \delta X^n g_{mn} \;,
\ee
with $m = 6,...,9$, so that we have an additional kinetic term $\frac{1}{2}\Tr (D\partial\delta X^n\partial \delta X^m g_{mn})$, resulting in
\be\label{final}
\begin{split}
S = -\frac{1}{g^2}\int d^5 x\sqrt{\det h}& \Big[
 \frac{1}{2}\partial_\mu\delta \Phi \partial^\mu \delta \Phi+\frac{1}{2\lambda^2} \partial_\mu\delta X^m\partial^\mu \delta X^m\\ &+\frac{1}{4}\left(\delta F_{\mu\nu}-  2  \mu \delta \Phi\omega_{\mu\nu}  \right) \left(\delta F^{\mu\nu}- 2 \mu \delta \Phi \omega^{\mu\nu} \right)\Big]\;.
\end{split}
\ee
 
Note that since we are dealing with a D4-brane  wrapping an $S^2$ there is a question about how to realise supersymmetry in the effective action. This requires twisting the theory by embedding the spin connection into a $\U(1)$ subgroup of the R-symmetry. We have five scalars $\delta \Phi , X^m$, the latter of which transform under a global $\SO(4)\simeq \SU(2)_A \times \SU(2)_B$. We choose to twist the $\U(1)_A\subset \SU(2)_A$. 

Usually, in the case of a one-complex-dimensional compactification, this twisting corresponds to wrapping the brane on a nontrivial supersymmetric 2-cycle, realised by a holomorphic curve. Even though here the $S^2$ is contractible, it is prevented from collapsing by the worldvolume flux and the supersymmetry twisting works the same way as in the topologically nontrivial cases \cite{Andrews:2006aw,Nastase:2009zu,Nastase:2010uy}. 

It is straightforward to turn the $\SO(4)$-invariant $\delta X^i$'s in terms of bosonic spinors on $S^2$. One can repackage them in terms of a new complex field
\be
q^{\dot \alpha} =\frac{ \sqrt 2}{\lambda} \doublet{X^6 + i X^8}{X^7 + i X^9}\ ,
\ee
through which we can rewrite the action as
\be\label{finalfinal}
S = -\frac{1}{g^2}\int d^5 x\sqrt{\det h} \Big[
 \frac{1}{2}\partial_\mu\delta \Phi \partial^\mu \delta \Phi +\frac{1}{4}\left(\delta F_{\mu\nu}-  2  \mu \delta \Phi \omega_{\mu\nu}  \right) \left(\delta F^{\mu\nu}- 2 \mu \delta \Phi \omega^{\mu\nu} \right)+\partial_\mu q^\dagger_{\dot\alpha}\partial^\mu q^{\dot\alpha}\Big]\;.
\ee
In the above one  can `pull out' a Hopf spinor from the transverse scalars by defining $q^{\dot\alpha} = Q^{\dot\alpha}_\alpha g^\alpha$. Then, following a large-$N$ version of the discussion in Sec.~5.3.3 of \cite{Nastase:2009zu,Nastase:2010uy}, \eqref{finalfinal}  contains exactly the  kinetic and mass terms for the bosonic T-spinors on the $S^2$, $\Xi$, as given in Eq.~(5.105) of that paper. 

Note that, when appropriately supersymmetrised as in \cite{Nastase:2009zu,Nastase:2010uy}, \eqref{finalfinal}  is the action for MSYM on $\mathbb R^{2,1}\times S^2$, despite the presence of the mass terms. Indeed, consider for instance  the case of ${\cal N}=4$ MSYM in four dimensions, arising on D3-branes. The theory on the $\mathbb R^2\times S^2$ space conformally equivalent to $\mathbb R^4$ has conformally coupled scalars, with the mass terms $-\frac{\mu^2}{2}\Phi^2$ coming from ${\cal R}\Phi^2$ (with ${\cal R}$ the Ricci scalar), and  mass terms for the other fields related to it by supersymmetry. For the T-dual D4-brane theory, the same thing happens.

It is straightforward to obtain the quadratic action for the full D4-brane fields from the fluctuation action \eqref{finalfinal}: One needs to replace $\partial_\mu\delta \Phi \partial^\mu \delta \Phi \rightarrow \partial_\mu \Phi \partial^\mu \Phi$ and $\delta F^{\mu\nu}- 2 \mu \delta \Phi\omega^{\mu\nu}\rightarrow F^{\mu\nu}-2\mu \Phi\omega^{\mu\nu}$, where $F^{\mu\nu}=\delta F^{\mu\nu}+F_0^{\mu\nu}$, $\Phi=\Phi_0+\delta \Phi$ and $F_0^{\mu\nu}=2\mu \Phi_0\omega^{\mu\nu}=2 \mu^2 N\omega^{\mu\nu}$ is the background solution. The full action thus obtained must admit the nonzero constant background $F_0$ and also be compatible with the twisted supersymmetry of the theory, as explained below Eq.~(\ref{final}).

\section{Higher order terms and non-Abelian generalisation}\label{nonabelian}

Until now we have only looked at quadratic Abelian fluctuations. However, it is useful to understand what happens to higher order terms, especially in view of generalising both the results of this paper and \cite{Nastase:2009ny,Nastase:2009zu,Nastase:2010uy} to the non-Abelian case. Due to gauge invariance, we expect that in the interacting theory the partial derivatives $\d_\mu$ are completed to covariant derivatives $D_\mu$. However, even then one does not expect to see $\delta \Phi^n$ interactions with $n> 2$. We will explicitly check this in the following from the point of view of the M2-brane theory fluctuations. One should in principle also compute other possible higher order terms involving different combinations of fields, but we will not attempt that here.

In order to proceed with the calculation, we should remind the reader some of the backdrop for \cite{Nastase:2009ny,Nastase:2009zu,Nastase:2010uy}: For the case of the mass-deformed ABJM theory \cite{Gomis:2008vc}, the ABJM scalars split as $Z^A = (R^\alpha, Q^{\dot\alpha})$, where $\alpha, \dot\alpha = 1, 2$. This reflects the breaking of the R-symmetry group $\SU(4)\to \SU(2)\times\SU(2)$. There is a set of zero-energy solutions where $Q^{\dot\alpha} = 0$. Then the equations of motion reduce to 
the vacuum equation (\ref{BPSvac})
\be\label{GRVV}
\frac{\mu k}{2 \pi}R^{\alpha} = R^{ \alpha} \Rd_{\beta} R^{ \beta} -   R^{ \beta} \Rd_{\beta} R^{ \alpha}\;.
\ee
The solutions are given by $R^\alpha = f G^\alpha$, where  $f=\sqrt{{\mu k}/{2 \pi}}$ and $G^\alpha$, $G^\dagger_\alpha$ are (anti)bi-fundamental $N \times N$ matrices. The single (Abelian) D4-brane is obtained by considering irreducible $G$'s satisfying \eqref{GRVV}, which were first given in \cite{Gomis:2008vc}. The fluctuations around these vacua can be organised according to 
\bea\label{ansatzflucts}
 R^{ \alpha } = f G^{\alpha} + r^{\alpha}\;, && 
 R^{\dagger}_{\alpha}  = f \Gd_{\alpha}  + r^{\dagger}_{\alpha}  \cr
 Q^{ \dot \alpha}  = q^{\dot \alpha }\;,\qquad \quad&& 
 Q^{\dagger}_{\dot \alpha}  = q_{\dot \alpha}^{\dagger}  \cr
 A_{\mu} = A_{\mu}\;,\qquad \quad &&  \psi^{ \dagger A } = \psi^{\dagger A }\;.
\eea
We will sketch how to extend these solutions and fluctuations in order to obtain a non-Abelian theory towards the end of this section. Relevant identities needed for the calculation, as well as a more precise definition of the continuum limit in which the matrices become functions on $S^2$, are given in the Appendix.

\subsection{Cubic and Quartic fluctuation action  from potential terms}\label{CQ}

Here we will extend the results of \cite{Nastase:2009ny,Nastase:2009zu,Nastase:2010uy} for the action of fluctuations in the mass-deformed ABJM theory by including cubic and quartic powers (as well as higher orders) of the `relative transverse' scalars. This corresponds to the radial scalar which we have been denoting as $\delta \Phi$ but, in order to keep with the conventions of that calculation, we will henceforth call simply  $\Phi$.

With the ABJM potential being sixth order, one could {\it a   priori} also get contributions to the fluctuation action coming from $\mathcal O(\Phi^5)$, $\mathcal O(\Phi^6)$ interactions. Since such terms cannot appear from the D4 MSYM action that we are comparing against, it is important to check that they vanish. We find that this is indeed the case.

The $r^\alpha$ fluctuations of \eqref{ansatzflucts} can be further decomposed at large $N$ in terms of\footnote{This follows from Eq.\ (4.76) of \cite{Nastase:2009ny}.}
\be\label{fluctuations}
r^\alpha = \frac{1}{2}\Phi G^\alpha +\frac{1}{2} K^i_a A_a (\tilde\sigma_i)^\alpha_\beta G^\beta\;,
\ee
where $(\tilde\sigma_i)^\alpha_\beta\equiv{(\tilde\sigma_i)^\alpha}_\beta$ are the transpose of the Pauli matrices, such that $[\tilde \sigma_i,\tilde\sigma_j] = -2i\epsilon_{ijk}\tilde\sigma_k $, and $K^a_i$ are components of Killing vectors on $S^2$, $a = \theta, \phi$. One of the four real degrees of freedom for the fluctuation $r^\alpha$ does not appear in the final D4 action since it plays the role of a Goldstone boson, eaten by the gauge field during the Higgs mechanism that renders the ABJM CS-gauge field dynamical \cite{Mukhi:2008ux}. 

It is also useful to define
\be
G^\alpha G^\dagger_\beta = J^\alpha_\beta\;, \qquad J_i = (\tilde \sigma_i)^\alpha_\beta  J^\beta_\alpha\;,\qquad J^\alpha_\beta = \frac{N-1}{2}\delta^\alpha_\beta +\frac{1}{2}J_i (\tilde\sigma_i)^\alpha_\beta\;,
\ee
with $[J_i, J_j] = 2i\epsilon_{ijk} J_k$,

The potential terms in the mass-deformed ABJM action relevant for cubic and quartic fluctuations are the sextic and quartic potential terms. The ABJM sextic potential 
\be
V_6 = \frac{4\pi^2}{3k^2} \sum_{i=1}^4 \hat V_i\;,
\ee
 is composed of 
\bea
\nn \hat V_1 &=&  - \Tr( Z^A \bar Z_A Z^B \bar Z_B Z^C \bar Z_C )\\
\nn \hat V_2 &=&  - \Tr ( \bar Z_A Z^A \bar Z_B Z^B \bar Z_C Z^C ) \\
 \nn \hat V_3 &=&  - 4 \Tr ( Z^{A} \bar Z_B Z^C \bar Z_A Z^B \bar Z_C ) \\
  \hat V_4 &=& ~ 6 \Tr ( Z^A \bar Z_B Z^B \bar Z_A Z^C \bar Z_C ) \;.
\eea 
The quartic potential of the mass-deformed theory is
\be
 V_4 =  \frac{ 8\pi \mu}{k} \Tr(R^{[\alpha}R^\dagger_\beta R^{\beta]}R^\dagger_\alpha)\;.
\ee

\subsubsection{The $\Phi^4$, $\Phi^5$ and $\Phi^6$ terms}

From the sextic potential terms we get:
\bea
\nn\hat V_1 &=& - \frac{15}{16} (N-1)^3 f^2 \Tr[\Phi^4]\\
\nn\hat V_2 &=& - \frac{3}{16}N f^2 \Tr[\Phi^2 J^\alpha_\beta \Phi^2 J^\beta_\alpha] -\frac{12}{16} f^2 \Tr[\Phi^2 J^\alpha_\beta \Phi J^\beta_\gamma \Phi J^\gamma_\alpha]\\
\nn\hat V_3 &=&  - \frac{12}{16}(N-2) f^2\Tr[\Phi^2 J^\alpha_\beta \Phi^2 J^\beta_\alpha] -\frac{12}{16} (N-1)^2 f^2 \Tr[\Phi^4]\\
\nn&&- \frac{48}{16}f^2\Tr[\Phi^2 J^\alpha_\beta \Phi J^\gamma_\alpha \Phi J^\beta_\gamma]\\
\nn \hat V^4 &=&  \frac{36}{16} (N-1) f^2\Tr[\Phi^2 J^\alpha_\beta \Phi^2 J^\beta_\alpha] + \frac{6}{16} N(N-1)^2 f^2 \Tr[\Phi^4]\\
&& +\frac{48}{16}(N-1) f^2\Tr[\Phi^3 J^\alpha_\beta \Phi J^\beta_\alpha] \;.
\eea

Summing the above and using $f^2 = \frac{\mu k}{2\pi}$
\begin{multline}
V_6 = \frac{\pi\mu}{8 k} \Big( - (3N-1)(N-1)^2\Tr[\Phi^4]  + (7N-4)\Tr[\Phi^2 J^\alpha_\beta \Phi^2 J^\beta_\alpha]  +16(N-1) \Tr[\Phi^3 J^\alpha_\beta \Phi J^\beta_\alpha]\\ - 4\Tr[\Phi^2 J^\alpha_\beta \Phi J^\beta_\gamma \Phi J^\gamma_\alpha] -  16\Tr[\Phi^2 J^\alpha_\beta \Phi J^\gamma_\alpha \Phi J^\beta_\gamma]\Big)\;.
\end{multline}
The contribution from the quartic potential is
\be
V_4 = \frac{\pi\mu}{8k} \Big(\Tr[2 \Phi^2 J^\beta_\alpha \Phi^2 J^\beta_\alpha] - 2 (N-1)^2 \Tr[\Phi^4] \Big)\;.
\ee 
The total contribution for $\mathcal O(\Phi^4)$ terms then is
\begin{multline}
V = \frac{\pi\mu}{8 k} \Big( - (3N+1)(N-1)^2\Tr[\Phi^4]  + (7N-2)\Tr[\Phi^2 J^\alpha_\beta \Phi^2 J^\beta_\alpha]  +16(N-1) \Tr[\Phi^3 J^\alpha_\beta \Phi J^\beta_\alpha]\\ - 4\Tr[\Phi^2 J^\alpha_\beta \Phi J^\beta_\gamma \Phi J^\gamma_\alpha] -  16\Tr[\Phi^2 J^\alpha_\beta \Phi J^\gamma_\alpha \Phi J^\beta_\gamma]\Big)\;.
\end{multline}

We next need to manipulate the above using the identities in the Appendix.
The potential contribution now becomes
\begin{multline}
V = \frac{\pi\mu}{8 k} \Big( -(N-1)(6N-11)\Tr[\Phi^4]  - \frac{3N-20}{4}\Tr[[J_i,\Phi^2] [J_i, \Phi^2] ]  -(N-1)\Tr[[J_i,\Phi^3] [J_i, \Phi]]\\ -3 i \epsilon_{ijk}\Tr[[J_i,\Phi^2] [J_j, \Phi] J_k \Phi]\Big)\;.
\end{multline}

But in the classical limit $[J_i,.]=-2iK_i^a\pd_a$ and $J_k=Nx_k$, and $\frac{1}{N}\Tr\rightarrow \frac{1}{4\pi}\int d^2\sigma\sqrt{\det \hat h}$,  so we get 
\be
V_{\Phi^4}=\frac{\mu}{32 k}\int d^2\sigma \sqrt{\det \hat h} \Big[-6N^3\Phi^4+3N^2(\pd_a\Phi^2)(\pd^a\Phi^2)+N^2(\pd_a\Phi^3)(\pd^a\Phi)\Big]\ ,
\ee
since the $\epsilon_{ijk}$ term is $\sim \omega^{ab}\pd_a\pd_b=0$. 

At this stage we need to note that in order for the action of fluctuations to result in an action on the $S^2$, one had to rescale the $A_a, \Phi$ and $Q$ fields by $\frac{1}{N}$ in the classical limit (Eq.~(6.5) of \cite{Nastase:2009ny}). This means that all the $\Phi^4$ terms evaluated above rescale to zero.

As a result, we do not need to separately calculate the $\Phi^5$ and $\Phi^6$ terms, which might have led to higher derivative terms for $\Phi$, since from the traces we can at most get $N^3$. Together with the $N$ one gets when converting $\Tr$ to $\int$ this becomes at most $N^4$, which means that after the rescaling these terms vanish, as they should. 

\subsubsection{The $\Phi^3$ terms}

We still need to check the contributions at order $\mathcal O(\Phi^3)$. From the sextic potential we get
\bea
\hat V_1&:& -\frac{20(N-1)^3}{8}f^3\Tr[\Phi^3]\cr
\hat V_2&:& -\frac{12N}{8}f^3\Tr[\Phi^2{J^\alpha}_\beta \Phi {J^\beta}_\alpha]-\frac{8f^3}{8}\Tr[\Phi{J^\alpha}_\beta\Phi {J^\beta}_\gamma\Phi {J^\gamma}_\alpha]\cr
\hat V_3&:& -\frac{4f^3}{8}\Big[12(N-1)^2\Tr[\Phi^3]+12(N-2)\Tr[\Phi^2{J^\alpha}_\beta \Phi {J^\beta}_\alpha]+ 8 \Phi J^\alpha_\beta\Phi J^\gamma_\alpha\Phi J^\beta_\gamma\Big]\cr
\hat V_4&:& \frac{6f^3}{8}\Big[4N(N-1)^2\Tr[\Phi^3]+16(N-1)\Tr[\Phi^2{J^\alpha}_\beta \Phi {J^\beta}_\alpha]\Big]\;.
\eea
Using the identities in the  Appendix, we get for the $\Phi^3$ terms in $V_6$
\bea
V_6&=&\sqrt{\frac{2\pi\mu}{k}}\frac{\mu}{24}\Big[60(N-1)\Tr[\Phi^3]-(6N-21)\Tr[[J_i,\Phi^2][J_i,\Phi]]\cr
&&-6i\epsilon_{ijk}\Tr[[J_i,\Phi] [J_j, \Phi] J_k \Phi]\Big]\;,
\eea
whereas for the $\Phi^3$ terms in $V_4$ we get
\bea
V_4&=&\frac{8\pi \mu}{k}\frac{f}{8}\frac{4}{2}\Big[\Tr[{J^\alpha}_\beta\Phi^2 {J^\beta}_\alpha \Phi]-(N-1)^2\Tr[\Phi^3]\Big]\cr
&=&\sqrt{\frac{2\pi\mu}{k}}\mu \Big[(N-1)\Tr[\Phi^3]+\frac{1}{4}\Tr[[J_i,\Phi^2][J_i,\Phi]]\Big]\;.
\eea
In total, we have for the $\Phi^3$ terms
\bea
V&=&\sqrt{\frac{2\pi\mu}{k}}\mu\Big[\frac{7(N-1)}{2}\Tr[\Phi^3]-\frac{N-27/6}{4}\Tr[[J_i,\Phi^2][J_i,\Phi]]\cr
&&-\frac{i}{4}\epsilon_{ijk}\Tr[[J_i,\Phi] [J_j, \Phi] J_k \Phi]\Big]\cr
&\rightarrow& \sqrt{\frac{2\pi\mu}{k}}\frac{\mu}{4\pi}\int d^2\sigma \sqrt{\det \hat h}\left[\frac{7N^2}{2}\Phi^3+N^2\pd_a\Phi^2\pd^a\Phi\right]\ ,
\eea
but this again rescales to zero.\footnote{Here it was essential that both $N^3$ and $N^2$ terms cancelled before the classical limit, corresponding to divergent and finite terms in the  limit.} 

In conclusion, the action for $\Phi$ (with all  other fields set to zero) is only quadratic. This is consistent with the fact that we expect there to be vacuum solutions of the form (\ref{VacSol}) to all orders. This in turn implies that any higher order power of $\Phi$ that appears in the action must also be accompanied by higher powers of $F_{ab}$. Therefore, if we consistently truncate to the two-derivative effective action then only quadratic powers of $\Phi$ should arise. However, we should note that this pertains  only to the Abelian MSYM action on a single D4-brane. One would have to separately check whether the above arguments also generalise to the full non-Abelian case.

\subsection{Non-Abelian Generalisation}

We now sketch how one can extend the agreement for the action of fluctuations to the case of multiple M5/D4-branes. This needs to be implemented both from the M2 and D4-brane perspectives and while allowing fluctuations up to quartic order.

From the M5/D4 side, the calculation is straightforward: One should use the non-Abelian form of the effective action \eqref{effective} and keep cubic and quartic orders in fluctuations. By gauge invariance the result should be a non-Abelian generalisation of \eqref{final} where the partial derivatives are replaced by covariant ones. On the other hand, the derivation of the fluctuation action from the ABJM side is somewhat more involved. We will next set this up in detail.

In the calculation of the Abelian theory, we have considered irreducible $G$'s satisfying \eqref{GRVV}.
 In order to obtain the action for fluctuations for a full non-Abelian D4-theory on the $S^2$ through the procedure of \cite{Nastase:2009ny} one needs to consider reducible representations, as in all matrix constructions of higher dimensional branes,\footnote{See   \eg\cite{KeskiVakkuri:1997ec}.} with each block independently satisfying \eqref{GRVV}. Of particular interest are the configurations which correspond to $m$ copies of equally sized $N \times N$ blocks, since in that case the D4-branes are coincident and one expects a worldvolume gauge symmetry enhancement to $\U(m)$.

The starting point for studying these configurations is to consider mass-deformed ABJM theory with gauge group $\U(Nm)\times \U(Nm)$ and the solutions
\be
G^\alpha_{Nm\times Nm} = G^\alpha \otimes \one_{m \times m}\;,\qquad {G^\dagger_\alpha}_{Nm \times Nm} = G^\dagger_\alpha \otimes \one_{m \times m}\;,
\ee
where the $G^\alpha, G^\dagger_\alpha$ are $N\times N$ matrices. Even though this might look like it is only going to describe  a collection of non-interacting spherical D4's, the full interacting non-Abelian theory can be obtained by allowing the fluctuations to take values in the whole $Nm \times Nm$ matrix. These will capture all the `open string' degrees of freedom, both on each as well as across different branes and can be expressed in terms of:
\bea\label{ansatz1}
 R^{ \alpha } = f G^{\alpha} T^0+ r^{\alpha}\;, && 
 R^{\dagger}_{\alpha}  = f \Gd_{\alpha} T^0  + r^{\dagger}_{\alpha}  \cr
 Q^{ \dot \alpha}  = q^{\dot \alpha }\;,\qquad \quad&& 
 Q^{\dagger}_{\dot \alpha}  = q_{\dot \alpha}^{\dagger}  \cr
 A_{\mu} = A_{\mu}\;,\qquad \quad &&  \psi^{ \dagger A } = \psi^{\dagger A }\;,
\eea
where $T^0 = \one_{m \times m}$ and \eg
\be\label{ansatz2}
r^\alpha =  r^{\alpha}_0 T^0 + r^{\alpha}_l T^l\;,
\ee
with $T^l$ a traceless generator of $\SU(m)$ and similar expansions for the rest of the fluctuating fields. 

It is then straightforward to see how the non-Abelian fields and interactions  will arise. The trace over the $Nm \times Nm$ matrices factorises over the fluctuations \eqref{ansatz1}-\eqref{ansatz2} as
\be
\Tr_{Nm \times Nm} \to \Tr_{N\times N} \Tr_{m \times m}\;.
\ee
In the large-$N$ limit this can be approximated by
\be
\Tr_{Nm\times Nm} \to \frac{N}{4\pi}\int d^2 \sigma \sqrt{\det \hat h}\; \Tr_{m \times m}\;,
\ee
with $\sigma = \theta, \phi$ and $\hat h$ the metric on the unit $S^2$.

 In this way the quadratic terms in the fluctuating fields of the mass-deformed ABJM theory trivially become adjoint fields in the $\U(m)$ gauge group for the D4-theory on $S^2$, as can be seen \eg for the $\pd_\mu \Phi \pd^\mu \Phi$ part of the $D_\mu \Phi D^\mu \Phi$ non-Abelian scalar kinetic term and similarly for all other fields. 

In order to obtain the full theory, involving gauge interactions coming from the covariant derivatives, one needs to also include cubic and quadratic fluctuations. This should be relatively straightforward, if not somewhat tedious. We have already seen that the Abelian parts of the $\Phi^3$ and $\Phi^4$ contributions are zero up to subleading terms in powers of $\frac{1}{N}$. Of course that does not exclude {\it a priori} terms of the type $f_{abc}\Phi^a\Phi^b\Phi^c$ or $[\Phi^a,\Phi^b]^2$, which have no Abelian component, and for which the off-diagonal-block fluctuation in the $Nm\times Nm$ matrix (corresponding to interactions between different branes) could give nonzero contributions. One would also need to obtain $[\Phi,A]^2$ as well as the $\pd \Phi [\Phi,A]$ terms in order to reproduce the full scalar kinetic term, as well as the equivalent contributions for the gauge fields.

\section{Momentum, Fluxes and Instantons}\label{fluxes}

In the ABJM description of M2-branes \cite{Aharony:2008ug} eleven-dimensional momentum modes are somewhat obscured. The reason for this is that  the natural action
\be
Z^A \to e^{i\theta}Z^A\ ,
\ee
is in fact a gauge rotation. Therefore it is not clear how to describe momentum modes along the $\U(1)$ that describes the common phases of the spacetime coordinates. However if we construct the Hamiltonian we find
\bea
\nonumber H&=& \int d^2 x\   \Tr (\Pi_{Z^A}\Pi_{\bar Z_A})+\Tr (D_i Z^A D^i\bar Z_A) +V\\
 &&+ \Tr  \left(iZ^A\Pi_{Z^A}-i\Pi_{\bar Z_A}\bar Z_A  - \frac{k}{2\pi}F^L_{12}\right)A^L_0\\
\nonumber&&+ \Tr  \left(i\bar Z_A\Pi_{\bar Z_A} -i\Pi_{Z^A}Z^A  + \frac{k}{2\pi}F^R_{12}\right)A^R_0\;,
\eea
where $V$ is the potential, $\Pi_{Z^A} = \partial_0 \bar Z_A$, and we have set the Fermions to zero for simplicity.
As is usual in a gauge theory, the timelike components of the gauge fields give rise to constraints. Thus we find 
\begin{eqnarray}\label{naflux}
\nonumber  \frac{k}{2\pi}F^L_{12} &=& iZ^A\Pi_{Z^A}-i\Pi_{\bar Z_A}\bar Z_A  \\
  \frac{k}{2\pi}F^R_{12}&=&i\Pi_{Z^A}Z^A-i\bar Z_A\Pi_{\bar Z_A}   \ .
\end{eqnarray}
In the case that the $Z^A$ are all diagonal with eigenvalues $z^A = \frac{1}{\sqrt2} e^{i\theta^A}$ one sees that 
\be
\frac{k}{2\pi}F^L_{12}  = \frac{k}{2\pi}F^R_{12}  = \sum_A \partial_0\theta^A\ .
\ee
Here we see that the `missing' eleven-dimensional momentum around the common $\U(1)$ phase is given by the magnetic flux, $F_{12}$. 

Let us now look at how the flux of the M2-brane worldvolume gets lifted to an instanton on the D4-brane.  In the following we will denote the non-Abelian $\U(N)$ gauge fields of the M2-brane worldvolume (after Higgsing) with a hat and the resulting D4 (M5) brane gauge fields without a hat. We note that according to the usual prescription of converting matrices to functions on the emergent two-sphere \cite{Nastase:2009ny} one has  the appearance of a relative normalisation factor along the $x^0,x^1,x^2$ directions:
\be
\frac{1}{N} \hat A^\mu_{N\times N}\to A^\mu  \;.
\ee

We next  need to consider the flux quantisation rule for $ A_\mu$, \ie the Abelian part of $\hat A_\mu$. In particular, note that in a $ \U(N)$ gauge theory we see that
\be
 \frac{1}{2\pi}\int_{{\mathbb R}^2}\hat F = Q \left(\begin{array}{ccc}
 1 &   &   \\
  & \ddots  &    \\
  &  &1     \\
  \end{array}\right) + \ldots\ ,
\ee
where the ellipsis denotes terms in the Lie algebra that involve the traceless generators of $\U(N)$ (\ie those of the $\SU(N)$ subalgebra). Thus we need to know the quantisation condition for the overall $\U(1)$ factor. To determine this we simply observe that a single $\U(1)$ generator can be written as
\be
\left(\begin{array}{ccc}
 1&   &   \\
  & 0  &    \\
  &  &\ddots     \\
  \end{array}\right) =\frac{1}{N} \left(\begin{array}{ccc}
 1 &   &   \\
  & \ddots  &    \\
  &  &1     \\
  \end{array}\right) + \ldots\ ,
\ee
where again the ellipsis denotes trace-free generators of the Lie algebra. Since the left hand side has the standard Dirac quantisation $2\pi {\mathbb Z}$ we conclude that  the identity flux component has charge quantisation $\frac{2\pi}{N} {\mathbb Z}$.\footnote{That is, if we allow for integer charges on the LHS then we must allow for fractional charges on the RHS.} More mathematically this fractional quantisation condition arises because $\U(N)\sim ( \U(1)\times \SU(N))/{\mathbb Z}_N$ as discussed for M2-branes in more detail in \cite{Lambert:2010ji}.
Thus we see that 
\be
Q = \frac{q}{N}\;,
\ee
with $q \in {\mathbb Z}$, and therefore 
\be
\frac{1}{2\pi N} \int_{{\mathbb R}^2}\Tr \hat F = \frac{q}{N} \qquad \to\qquad  \frac{1}{2\pi} \int_{{\mathbb R}^2} F  = \frac{q}{N}\ ,
\ee
with $q \in {\mathbb Z}$.

Finally we remind that the D4 (M5) brane configuration includes the background flux \eqref{backflux}:
\be
\frac{1}{ 2\pi }\int_{S^2} F =N\ .
\ee
With these ingredients we see that
  the instanton number is
\be
\frac{1}{8\pi^2} \int_{{\mathbb R}^2\times S^2} F\wedge F=\frac{1}{4\pi^2} \int_{{\mathbb R}^2} F\int_{S^2} F = q \in {\mathbb Z}\;.
\ee
The D4-brane action on $\mathbb R^{2,1}\times S^2$ has therefore  states carrying nonzero  instanton number equal to an arbitrary integer.

We can extend this argument to  the non-Abelian case corresponding to $m$ D4(M5)-branes. Here the D4(M5)-brane background gauge field is
\be
\frac{1}{ 2\pi }\int_{S^2} F =\left(\begin{array}{ccc}
 N_1&   &   \\
  & \ddots &    \\
  &  &N_m     \\
  \end{array}\right) \;.
\ee
To turn on momentum in ABJM we need to consider fluxes of the form
\be
\frac{1}{ 2\pi }\int_{{\mathbb R}^2} F =\left(\begin{array}{ccc}
 Q_1&   &   \\
  & \ddots &    \\
  &  &Q_m    \\
  \end{array}\right) \;.
\ee
Such a flux arises from the reducible M2-brane flux
\be
\frac{1}{ 2\pi }\int_{{\mathbb R}^2} \hat F = \left(\begin{array}{ccc}
 Q_1{\mathbb I}_{N_1\times N_1} &   &   \\
  & \ddots &    \\
  &  &Q_m  {\mathbb I}_{N_m\times N_m}   \\
  \end{array}\right) \;.
\ee
To determine the quantisation rule for $Q_i$ we observe that
\be
\left(\begin{array}{ccc}
 Q_1{\mathbb I}_{N_1\times N_1} &   &   \\
  & 0  &    \\
  &  &\ddots     \\
  \end{array}\right) =\frac{Q_1N_1}{N_1+\ldots + N_m} \left(\begin{array}{ccc}
 1 &   &   \\
  & \ddots  &    \\
  &  &1     \\
  \end{array}\right) + \ldots\ ,
\ee
where again the ellipsis denotes trace-free terms. Since the coefficient on the right hand side must be of the form $q/(N_1+\ldots + N_m)$ we obtain that 
\be
Q_i = \frac{q_i }{N_i}\;, \qquad q_i\in {\mathbb Z}\;.
\ee
From these we deduce that the instanton number is
\be
\frac{1}{8\pi^2}\Tr  \int_{{\mathbb R}^2\times S^2} F\wedge F=\frac{1}{4\pi^2} \Tr \int_{{\mathbb R}^2} F\int_{S^2} F = \sum_i Q_iN_i  = q_1+\ldots + q_m\;.
\ee
Thus we see that the ABJM prescription for momentum through the eleventh dimension, given by magnetic flux,  is precisely mapped into the instanton
number in the D4-brane description.

It is important to emphasise that the instanton states that we are referring to here are not necessarily the usual selfdual solutions but any state in the 5D MSYM theory which carries nonzero $\int F\wedge F$ as a result of the fluxes.  Indeed, one can see from the discussion at the end of Section~\ref{overalltransverse} that the on-shell action for the configuration with $F_{12}\neq 0$ and $F^{ab}=F_0^{ab}$ (the only one allowed by the equations of motion), receives a nonzero contribution just from $\int F_{12}^2$. Therefore the instanton number $\sim \int F_{12}\wedge F_{\theta_1\phi_1}$ does not appear as the usual topological contribution to the on-shell action. One could of course also find customary instanton configurations, particularly in the non-Abelian case, where the dynamics of the sphere directions do contribute to the action. These would involve turning on nontrivial scalar fields while still having  an on-shell action quantised in terms of the instanton number.

\subsection{M5-brane picture}

 In Section~\ref{analysis} we mapped the action for fluctuations of the M2-brane action, including $F_{12}$ flux, to those of the D4-brane theory on $\mathbb R^{2,1} \times S^2$. Moreover, we argued that turning on this flux corresponds to turning on units of momentum around the M-theory circle. We now finally show that this is compatible with the expected spacetime interpretation of an  M5-brane wrapping $S^3/\mathbb Z_k$.

Let us consider the case of $N$ M2-branes expanding into a single M5. For concreteness suppose that the spatial dimensions $x^1,x^2$ are compactified on a torus or size $L$. Without turning on any additional worldvolume fluxes the action is
\be
 S = -M=-\frac{\lambda N}{2} T_{D4} 4\pi L^2 = -\frac{L^2N}{4\pi^2 l_p^3}\;,
\ee
where we have used once again that $T_{D4}=(2\pi)^{-4}l_p^{-6}R_*$, $\lambda = 2\pi l_s^2$, $R_*=g_sl_s$ and $l^3_p=g_sl_s^3$. This corresponds to an M5-brane wrapped on $T^2\times S^3$  (or a D4-brane wrapped on $T^2\times S^2$) in the presence of background four-form flux. 

 Let us now include the effect of worldvolume flux.  According to our discussion above, the allowed flux that corresponds to turning on $q$ units of eleven-dimensional momentum is
\be
\delta F_{12} = \frac{2\pi q}{L^2 N}\;,\qquad q\in {\mathbb Z}\ .
\ee
The action becomes
\bea
\nonumber
S &=& -M\left(1+\frac{1}{2}\frac{R_0^2}{4\mu^2N^2}\delta F_{12}\delta F_{12}\right) \\
\nonumber &=& -M\left(1+\frac{1}{2}\frac{4\pi^2 q^2R_0^2}{4\mu^2N^4L^4}
\right)\\
\nonumber  &=& -M\left(1+\frac{1}{2M^2}\frac{R_0^4}{16\pi^2\mu^2N^2l_p^6}\left(\frac{q}{R_0}\right)^2
\right)\\
 &=& -\left(M + \frac{1}{2M}\left(\frac{kq}{R_0}\right)^2\right)\;.
\eea
This precisely agrees with the action of a single M5-brane wrapped on $T^2\times S^3/{\mathbb Z}_k$ that carries momentum $\frac{q}{R_*}$, such that the action is $S = -\sqrt{M^2+(\frac{q}{R_*})^2}$ when expanded to second order in $q$.

Finally let us comment on the extra term that appears in the D4-brane analysis of the fluctuations that was mentioned in footnote 5. This term gives rise to a Chern-Simons like coupling on the five-dimensional Yang-Mills theory of the D4-brane: $ \omega\wedge d\delta F\wedge \delta A$. If we include this term then one finds that solutions with non-zero instanton number are excluded. This may seem paradoxical, however we note that it is derived from the D4-brane effective action which, by construction, is not valid when there is non-vanishing eleven-dimensional momentum. Therefore the appearance of this term is consistent with the D4-brane analysis. On the other hand,  as we have argued above, there is no obstruction to turning on magnetic flux in the ABJM theory and indeed  this Chern-Simons term appears to be absent from the five-dimensional Yang-Mills effective action obtained from M2-branes \cite{Nastase:2009ny}. It would be interesting to reconcile this observation with the recent results of \cite{Gustavsson:2011af,Linander:2011jy} and the role of supersymmetry.

\section{Conclusions}\label{conclusions}

In this paper we have studied M2-branes in a background four-form flux. The resulting system expands via the Myers effect into M5-branes wrapped on a fuzzy $S^3$. We computed the effective action of a static M5-brane in this background and showed that it stabilised at the same radius as predicted by the M2-brane gauge theory. In addition, by reducing to type IIA string theory, we derived the fluctuation action of the associated D4-brane wrapped on $S^2$. These are determined by five-dimensional MSYM and also agree with the fluctuations about the M2-brane vacuum that were obtained in \cite{Nastase:2009ny}.

We next considered the effect of introducing  worldvolume magnetic flux into the  worldvolume description of M2-branes. According to  ABJM  \cite{Aharony:2008ug} this corresponds to introducing  momentum along the eleventh-dimension of M-theory. We showed that this was equivalent to introducing instanton flux in the five-dimensional  MSYM description of the D4-brane theory.  Since the ABJM description captures the full M-theory dynamics of M2-branes, we thus conclude that the five-dimensional MSYM theory on $\mathbb R^{2,1}\times S^2$, Eq.~\eqref{finalfinal}, when one includes all states carrying nonzero instanton charge, captures the full M5-brane degrees of freedom on $\mathbb R^{2,1}\times S^3/{\mathbb Z}_k$. This is in agreement with the conjecture of \cite{Douglas:2010iu,Lambert:2010iw}.

\section*{Acknowledgments}

The research of HN is supported in part by CNPQ grant 301219/2010-9. CP would like to acknowledge support from the DOE under grant DE-FG02-96ER40959 and thank the CERN Theory Division for hospitality during the course of this work.

\begin{appendix}

\section{Identities and formalism needed for cubic/quartic fluctuations}\label{A}

In this Appendix we gather identities used in the calculation of $\Phi^3,\Phi^4$ terms in the action for fluctuations. Most of these can be 
found in \cite{Nastase:2009ny,Nastase:2009zu,Nastase:2010uy}.

For $G^\a$ and $G^\dagger_\a$, we have the identities
\bea
G^\alpha G^\dagger_\alpha= J = N-1\qquad&&\qquad G^\dagger_\alpha G^\alpha = \bar J = N(1 - E_{11})\\
G^\alpha \bar J = N G^\alpha \qquad &&\qquad \bar J G^\dagger_\alpha  = N G^\dagger_\alpha\\
J^\alpha_\beta J^\gamma_\alpha = (N-2)J^\gamma_\beta  + \delta^\gamma_\beta J \qquad&&\qquad J^\alpha_\beta J^\beta_\alpha = N J\;.
\eea

 In the continuum limit one can identify $[J_i , \cdot ] = -2i \epsilon_{ijk}x_j \pd_k = -2 i K_i^a \pd_a$ and $x_i = \frac{J_i}{\sqrt{N^2-1}}$, as well as $\Tr \to \frac{N}{4\pi} \int d\sigma^2 \sqrt {\det \hat h}$, where $\hat h$ is the dimensionless unit metric on $S^2$, and the matrix fluctuations become fields on the sphere. 

Definitions and some identities for $K_i^a$ follow:
\bea
K_1^\theta = -\sin{\phi} && K_1^\phi = -\cot{\theta}\cos{\phi}\cr
K_2^\theta = \cos \phi~~~  && K_2^\phi = -\cot{\theta}\sin{\phi}\cr
K_3^\theta = 0~ ~~~~~~~&& K_3^\phi = 1\;.
\eea
 The relations between Cartesian and spherical  coordinates is
\bea
x_1 &=& \sin{\theta}\cos{\phi} \cr
x_2 &=& \sin{\theta}\sin{\phi}\cr
x_3 &=& \cos{\theta}\;.
\eea
One can then explicitly evaluate the sets of identities 
\bea
K_i^aK_i^b&=&\hat h^{ab}\cr
\epsilon_{ijk} x_i K_j^a K_k^b&=&\hat\omega^{ab} =\frac{\epsilon^{ab}}{\sqrt{\hat h}} \cr
 K_i^ah_{ab}K^b_j&=&\delta_{ij}-x_ix_j\cr
K_i^a\pd_a K_i^b&=&\frac{1}{\sqrt{\hat h}}\pd^b\sqrt{\hat h}\;.
\eea
Further identities that were used for calculations include
\bea
&& x_i \pd^a K_i^b=\hat\omega^{ab}\cr
&&\epsilon_{ijk}\pd_a K_i^b x_j K_k^a=0\cr
&& \epsilon_{ijk}\pd_a K_i^b K_j^c K_k^a\times (sym. b\leftrightarrow c)=0\cr
&& (\pd_a x_i)K_j^a=\epsilon_{ijk}x_k\;.
\eea
From the last relation we also obtain 
\bea
 (\pd_a x_i)K_i^a&=&0\cr
 \epsilon_{ijk}(\pd_a x_i)K_j^a x_k&=&2\cr
 \epsilon_{ijk}(\pd_a x_i) K_j^a K_k^b&=&0\;.
\eea

Some useful identities for objects appearing in the sixth order scalar potential are
\bea
\Tr[J^\alpha_\beta A J^\beta_\alpha B] &=& \frac{(N-1)^2}{2}\Tr[AB] +\frac{1}{2}\Tr[A J_i B J_i] \nn\\
&=& N(N-1) \Tr [AB]+ \frac{1}{4}\Tr[[J_i, A][J_i, B]] \\
\Tr[A J^\alpha_\beta B J^\beta_\gamma C J^\gamma_\alpha] &=&\frac{(N-1)^3}{4} \Tr[ABC] + \frac{N-1}{4}\Tr[ABJ_i C J_i+ BC J_i A J_i +CA J_i B J_i]\nn\\ &&-\frac{i}{4}\epsilon_{ijk}\Tr[AJ_i B J_j C J_k]\\
\Tr[A J^\alpha_\beta B J^\gamma_\alpha C J^\beta_\gamma] &=&\frac{(N-1)^3}{4} \Tr[ABC] + \frac{N-1}{4}\Tr[ABJ_i C J_i+ BC J_i A J_i +CA J_i B J_i]\nn\\ &&+\frac{i}{4}\epsilon_{ijk}\Tr[AJ_i B J_j C J_k] \;.
\eea
In our case these translate into
\bea
\Tr[J^\alpha_\beta \Phi^2 J^\beta_\alpha \Phi^2] &=& \frac{(N-1)^2}{2}\Tr[\Phi^4] +\frac{1}{2}\Tr[\Phi^2 J_i \Phi^2 J_i]\nn \\
&=& N(N-1) \Tr [ \Phi^4]+ \frac{1}{4}\Tr[[J_i,  \Phi^2][J_i,  \Phi^2]] \\
\Tr[J^\alpha_\beta  \Phi^3 J^\beta_\alpha  \Phi] &=& \frac{(N-1)^2}{2}\Tr[ \Phi^4] +\frac{1}{2}\Tr[ \Phi^3 J_i  \Phi J_i] \nn\\
&=& N(N-1) \Tr [ \Phi^4]+ \frac{1}{4}\Tr[[J_i,  \Phi^3][J_i,  \Phi]] \\
\Tr[ \Phi^2 J^\alpha_\beta  \Phi J^\beta_\gamma  \Phi J^\gamma_\alpha] &=&\frac{(N-1)^3}{4} \Tr[ \Phi^4] + \frac{N-1}{4}\Tr[2 \Phi^3J_i  \Phi J_i+  \Phi^2 J_i  \Phi^2 J_i]\nn\\ 
&&-\frac{i}{4}\epsilon_{ijk}\Tr[ \Phi^2 J_i  \Phi J_j  \Phi J_k]\\
&=&\frac{(N-1)^2(2N+1)}{2}\Tr[\Phi^4]+\frac{N-1}{8}\Tr[2[J_i,\Phi][J_i,\Phi^3]+[J_i,\Phi^2][J_i,\Phi^2]]\cr
&&-\frac{i}{4}\epsilon_{ijk}\Tr[ \Phi^2 J_i  \Phi J_j  \Phi J_k]\\
\Tr[ \Phi^2 J^\alpha_\beta  \Phi J^\gamma_\alpha  \Phi J^\beta_\gamma] &=&\frac{(N-1)^3}{4} \Tr[ \Phi^4] + \frac{N-1}{4}\Tr[2 \Phi^3J_i  \Phi J_i+  \Phi^2 J_i  \Phi^2 J_i]\nn\\ &&+\frac{i}{4}\epsilon_{ijk}\Tr[ \Phi^2 J_i  \Phi J_j  \Phi J_k] \cr
&=&\frac{(N-1)^2(2N+1)}{2}\Tr[\Phi^4]+\frac{N-1}{8}\Tr[2[J_i,\Phi][J_i,\Phi^3]+[J_i,\Phi^2][J_i,\Phi^2]]\cr
&&+\frac{i}{4}\epsilon_{ijk}\Tr[ \Phi^2 J_i  \Phi J_j  \Phi J_k]\ .
\eea
We also have 
\bea
\epsilon_{ijk}\Tr[J_i\Phi J_j \Phi J_k \Phi^2]&=&\epsilon_{ijk}\Tr[[J_i,\Phi] [J_j, \Phi] J_k \Phi^2]+2i (N^2-1)\Tr[\Phi^4]\cr
&&+i \Tr[[J_i,\Phi^2][J_i,\Phi^2]]\ ,
\eea
as well as
\bea
&&\Tr[\Phi^2{J^\alpha}_\beta \Phi{J^\beta}_\alpha]=N(N-1)\Tr[\Phi^3]+\frac{1}{4}\Tr[[J_i,\Phi^2][J_i,\Phi]]\\
&&\Tr[\Phi{J^\alpha}_\beta\Phi{J^\beta}_\gamma\Phi{J^\gamma}_\alpha]=N^2(N-1)\Tr[\Phi^3]+\frac{3N-1}{8}\Tr[[J_i,\Phi^2][J_i,\Phi]]\cr
&&-\frac{i}{4}\epsilon_{ijk}\Tr[[J_i,\Phi] [J_j, \Phi] J_k \Phi]\\
&&\Tr[\Phi{J^\alpha}_\beta\Phi{J^\gamma}_\alpha\Phi{J^\beta}_\gamma]=(N^2-N-1)(N-1)\Tr[\Phi^3]+\frac{3N-5}{8}\Tr[[J_i,\Phi^2][J_i,\Phi]]\cr
&&+\frac{i}{4}\epsilon_{ijk}\Tr[[J_i,\Phi] [J_j, \Phi] J_k \Phi]\;.
\eea

\end{appendix}

\bibliographystyle{utphys}
\bibliography{M2M5}

\end{document}